# Epitaxial growth of a two-dimensional topological insulator candidate: monolayer $Si_2Te_2$

*Xiaochun Huang[1,†,\*], Rui Xiong[2,†], Klara Volckaert[3], Chunxue Hao[1], Deepnarayan Biswas[3], Marco Bianchi[3], Philip Hofmann[3], Philip Beck[1], Jonas Warmuth[1], Baisheng Sa[2]\*, Jens Wiebe[1]\*, and Roland Wiesendanger[1]\**

[1]Department of Physics, University of Hamburg, Hamburg, Germany.

[2]Multiscale Computational Materials Facility, Key Laboratory of Eco-materials Advanced Technology, College of Materials Science and Engineering, Fuzhou University, Fuzhou, PR China.

[3]Department of Physics and Astronomy, Interdisciplinary Nanoscience Center (iNANO), Aarhus University, 8000 Aarhus C, Denmark.

\*Corresponding author. Email: xhuang@physnet.uni-hamburg.de; bssa@fzu.edu.cn; jwiebe@physnet.uni-hamburg.de; rwiesend@physnet.uni-hamburg.de

†These authors contributed equally to this work.

**Hexagonal Si$_2$Te$_2$ monolayers (ML-Si$_2$Te$_2$) were predicted to show strain-dependent band-crossover between semiconducting and room-temperature quantum spin Hall phases. However, investigations on this artificial two-dimensional (2D) material have mainly been restricted to theoretical calculations because its bulk counterpart does not exist naturally. Here, we report on the successful epitaxial growth of ML-Si$_2$Te$_2$ films on Sb$_2$Te$_3$ thin film substrates. High-quality (1×1) ML-Si$_2$Te$_2$ films with a coverage as high as 95% were obtained as revealed by scanning tunneling microscopy. X-ray photoelectron spectroscopy confirms the absence of intermixing between Si$_2$Te$_2$ and Sb$_2$Te$_3$ at the interface. By combining scanning tunneling spectroscopy with density functional theory calculations, we demonstrate the semiconducting band structure of ML-Si$_2$Te$_2$ on Sb$_2$Te$_3$. Furthermore, it is theoretically predicted that the system can be driven into the nontrivial phase via reducing the strain by 4.4% using strain engineering. Our results pave the way for in-depth investigations on this 2D topological insulator candidate.**

## INTRODUCTION

Two-dimensional topological insulators (2D TIs), namely, quantum spin Hall (QSH) insulators, have an insulating bulk but dissipationless one-dimensional transport edge channels, which are expected to be used for applications in spintronics and quantum computers.[1-7] Since the experimental confirmation of the QSH state in HgTe/CdTe and InAs/GaSb quantum wells,[4,5] intensive research interest exists in uncovering new 2D TI materials. Thus far, a variety of 2D TIs, such as stanine,[8] Bi bilayers,[9,10] bismuthene on SiC,[11] ZrTe$_5$/HfTe$_5$,[12,13] and WTe$_2$ monolayers,[14-16] have been proposed theoretically and confirmed experimentally. In view of practical applications of QSH-based devices, searching for 2D TIs with sizable band gap is of significant importance. Especially, considering the compatibility with modern semiconductor technology, Si-based candidates are most desirable. For this reason, a lot of efforts were devoted to the investigation of silicene.[17-19] Unfortunately, the strong hybridization between silicene and silver substrates obstructs practical applications.

Recently, a room-temperature QSH phase was predicted in an artificial 2D material, monolayer $Si_2Te_2$ (ML-$Si_2Te_2$), when its lattice constant resides within a specific range around the free-standing case.[20] This Si-based 2D material has hexagonal lattice (*P-3m1*) symmetry with a unique Te-Si-Si-Te stacking sequence, and its band topology can be tuned by reasonable strain-engineering.[21] Although the structural stability of ML-$Si_2Te_2$ has been confirmed by density functional theory (DFT) calculations from different groups,[20,21] bulk $Si_2Te_2$ does not exist naturally, prohibiting the fabrication of ML-$Si_2Te_2$ from its 3D counterpart.[22,23] Thus, the critical question is whether it is possible to experimentally synthesize ML-$Si_2Te_2$ films with a clean surface for further studies. Although $Si_2Te_2$ was supposedly synthesized within a sandwich heterostructure,[24,25] the experimental characterization of the surface morphology and the electronic properties of ML-$Si_2Te_2$ remains challenging. In this work, we realize the epitaxial growth of ML-$Si_2Te_2$ films on $Sb_2Te_3$ thin film substrates by molecular beam epitaxy (MBE). Using scanning tunneling microscopy (STM), we reveal a (1×1) lattice of the obtained homogeneous ML-$Si_2Te_2$ films. X-ray photoelectron spectroscopy (XPS) confirms the existence of Si-Te bonds in this new material and the absence of intermixing between $Si_2Te_2$ and $Sb_2Te_3$ at the interface. To unambiguously characterize its electronic structure around the Fermi level, we performed scanning tunneling spectroscopy (STS) measurements on a sample of ML-$Si_2Te_2$ grown on a semiconducting one quintuple layer (QL) thin $Sb_2Te_3$ film. By comparing the measured explicit band gap with DFT calculations, we reveal the semiconducting band structure of ML-$Si_2Te_2$ on $Sb_2Te_3$. Besides, based on our results, we propose that strain engineering will enable the growth of ML-$Si_2Te_2$ in the QSH phase.

**RESULTS AND DISCUSSIONS**

Figures 1a,b schematically shows a heterostructure of ML-Si$_2$Te$_2$ on 1QL-Sb$_2$Te$_3$, which serves as a model-type configuration for our DFT calculations throughout this work (details are explained in the Supporting Information). The lattice vectors of the triangular lattice of Si$_2$Te$_2$ are labeled by the red arrows in Figure 1a, and the distance marked in Figure 1b is derived from DFT calculations. The samples were grown in a home-built ultrahigh vacuum (UHV) MBE chamber (base pressure < 3.5 × 10$^{-10}$ Torr).[26] A well-developed two-step procedure[27] was used to prepare high-quality Sb$_2$Te$_3$ films by co-evaporating high-purity Te and Sb onto a graphene/4H-SiC(0001) substrate (for more details, please see Supporting Information section S1). Monolayer Si$_2$Te$_2$ films were then prepared by co-evaporating Si and Te onto the Sb$_2$Te$_3$ surface (at 185 °C), followed by a post-annealing process (at 310 °C) under Te flux. In Figure S2, a series of STM images is presented showing the evolution of the surface morphology of ML-Si$_2$Te$_2$ while increasing the annealing temperature from 185 °C to 310 °C. We confirmed that high-quality Si$_2$Te$_2$ films were obtained only if the annealing temperature was raised to 310 °C, and also that Si atoms are neither doped into nor react with the Sb$_2$Te$_3$ film (see Supporting Information section S2 and S3). Figure 1c presents a typical STM image of an as-grown sample with a Sb$_2$Te$_3$ film thickness of >18 QL, showing that almost the whole Sb$_2$Te$_3$ surface is covered by a continuous epitaxial film, which is only interrupted by steps of the underlying Sb$_2$Te$_3$ (see Figure 1d). Small areas of the exposed Sb$_2$Te$_3$ film (marked by white arrows in Figure 1c) enable us to measure the step height of the epitaxial film on Sb$_2$Te$_3$. The apparent height extracted from the line profile is (0.65 ± 0.06) nm (see Figure 1d), close to the calculated spacing (0.74 nm) between the surfaces of ML-Si$_2$Te$_2$ and Sb$_2$Te$_3$ (see Figure 1b and Supporting Information section S2). An atomic-resolution STM image shows the triangular arrangement of surface atoms of both the epitaxial film and the Sb$_2$Te$_3$ substrate simultaneously (see Figure 1e). The white dashed lines indicate that the ML-Si$_2$Te$_2$ lattice orientation is aligned to that of the Sb$_2$Te$_3$ substrate. The hexagonal symmetry of the epitaxial film can be further demonstrated by the fast Fourier transform of an atomic-resolution STM image of a defect-free

area (see Figure 1f). We note that the experimentally determined lattice constant ($a_{exp}$ = 4.15 Å) is in very good agreement with the theoretically predicted value of $a_{DFT}$ = 4.13 Å of ML-$Si_2Te_2$ derived from DFT calculations, indicating that a (1×1) ML-$Si_2Te_2$ film has been obtained without any type of reconstruction. As shown by Figure 1e, the $Si_2Te_2$ monolayer can easily be distinguished from the $Sb_2Te_3$ film by the bright and dark patches, which result from a random apparent height modulation on the picometer scale (±0.06 nm) as extracted from the line profile in Figure 1d. Despite this apparent height modulation, which is also visible in Figure 1f, the $Si_2Te_2$ monolayer film exhibits a continuous atomic lattice structure across the bright and dark patches. Considering the different lattice constants between free-standing ML-$Si_2Te_2$ (3.88 Å)[20] and bulk-$Sb_2Te_3$ (4.26 Å),[28] we suppose that these patterns are induced by interfacial strain. Similar features in STM images have been obtained on other epitaxial 2D materials and were interpreted in terms of interfacial strain effects as well.[8,29,30] Note that, in the Si-Te phase diagram, there is only one binary phase, $Si_2Te_3$, which has a much larger lattice constant ($a$ = 7.430(5) Å) compared with our experimentally determined value[22,23]. Besides, several theoretically predicted phases of the Si-Te system have been considered and excluded based on their lattice and electronic structures (see Supporting Information section S2).

To elucidate the elemental composition and chemical bonding states, we performed XPS measurements on ML-$Si_2Te_2$/thick-$Sb_2Te_3$ films and the bare $Sb_2Te_3$ substrate for comparison. Large-scale STM images were taken in advance and are shown in Figure 2a. The high quality of both samples is evident. The monolayer $Si_2Te_2$ film covers about 95% of the underlying $Sb_2Te_3$ thick film (see lower left panel), which is a good starting point for the subsequent XPS measurements. For the bare $Sb_2Te_3$ substrate, as shown in Figure 2b,c, as expected, both Te-4$d$ and Sb-4$d$ core-level spectra reveal single-component doublets. The corresponding binding energies of Te-4$d_{3/2}$ (41.26 eV) and Te-4$d_{5/2}$ (39.79 eV), and of Sb-4$d_{3/2}$ (33.75 eV) and Sb-4$d_{5/2}$ (32.52 eV) are consistent with previously reported values.[31] After the growth of $Si_2Te_2$, the XPS signals from the $Sb_2Te_3$ substrate are still detectable, which manifest themselves by their peak positions, Te-I-4$d_{3/2}$ (41.31 eV), Te-I-4$d_{5/2}$ (39.83 eV), Sb-4$d_{3/2}$ (33.76 eV), and Sb-4$d_{5/2}$ (32.51 eV), as shown in Figure 2e,f. Note that the unambiguously observed single-component doublet of Sb-4$d$ indicates that the Sb$_2$Te$_3$ substrate is not intermixing with Si$_2$Te$_2$. By comparison with Figure 2c, the broadened Sb-4d peaks in Figure 2(f) were attributed to the different surface chemical environment between the bare Sb$_2$Te$_3$ substrate and the Si$_2$Te$_2$/Sb$_2$Te$_3$ sample. As shown in Figure 2d,e, the Si-Te bond in Si$_2$Te$_2$ gives rise to the Si-2$p_{1/2}$ (at 100.62 eV) and Si-2$p_{3/2}$ (at 100.01 eV) peaks, and furthermore to the emergence of a new Te-4$d$ component, Te-II-4$d_{3/2}$ (41.77 eV) and Te-II-4$d_{5/2}$ (40.28 eV). Our XPS results, therefore, confirm the Si-Te bond in this new material Si$_2$Te$_2$ and highlight the absence of intermixing at the interface between Si$_2$Te$_2$ and Sb$_2$Te$_3$.

By combining STS with DFT calculations, further evidence for the realization of ML-Si$_2$Te$_2$ films was obtained from electronic structure investigations. In Figure 3a, we show representative d$I$/d$V$ spectra measured on ML-Si$_2$Te$_2$ and on the exposed Sb$_2$Te$_3$ thick-film substrate (>15QL). The spectrum of ML-Si$_2$Te$_2$ (red curve) can easily be distinguished from that of Sb$_2$Te$_3$ (blue curve) by four prominent peaks (marked by black arrows) below the Fermi level. As presented in Figure 3b, these peak features are qualitatively captured by the DFT calculations. An overall energy shift of the measured peak positions, compared to the calculated local density of states (LDOS), may result from a charge-transfer into the ML-Si$_2$Te$_2$ by the well-known $p$-doping effect of the Sb$_2$Te$_3$ substrate.[27,32] Different thicknesses of Sb$_2$Te$_3$ films induce slightly different $p$-doping in ML-Si$_2$Te$_2$, which is not included in the calculations. The d$I$/d$V$ spectra measured on ML-Si$_2$Te$_2$ at different sample spots show consistent results regarding the energy positions of those peaks (see Supporting Information section S6). Figure 3c,d shows the close-up views of the d$I$/d$V$ spectra and the calculated LDOS. The measured spectra of Sb$_2$Te$_3$ are consistent with previous studies.[27] The expected V-shape segment and a local minimum of the differential tunneling conductance, marked by a green arrow in the inset of Figure 3c, are detected and ascribed to the topological surface state (TSS) and the associated

Dirac point of $Sb_2Te_3$.[27,32,33] Interestingly, a very similar behavior of the electronic states is also observed for spectra taken on ML-$Si_2Te_2$ (see the red curve in Figure 3c), in the energy range from -0.1 V to +0.35 V. This can reasonably be understood based on Figure 3d. As shown, the calculated DOS predicts a bandgap (230 meV) of ML-$Si_2Te_2$, which means ML-$Si_2Te_2$ would electronically behave as a barrier layer with the Fermi energy within the bandgap. Thus, the observed V-shape segment and related features (within the gap) in d$I$/d$V$ spectra of ML-$Si_2Te_2$ are not arising from its intrinsic states but from the presence of the $Sb_2Te_3$ substrate. As a benefit of the weak Van der Waals (vdW) interaction between ML-$Si_2Te_2$ and $Sb_2Te_3$ (see Supporting Information section S7), these states are not significantly modified by the interfacial interaction. This conclusion is confirmed, in the following, by d$I$/d$V$ spectra taken on ML-$Si_2Te_2$ film grown on a 1QL-$Sb_2Te_3$ substrate.

It is well known that the critical thickness of the topological phase of $Sb_2Te_3$ is 4QL, which means that the TSS of $Sb_2Te_3$ opens a hybridization gap for thicknesses below 4QL.[32,33] Especially at 1QL thickness, the lower branch of the TSS is absent, leading to a considerable band gap (660 meV).[33] This provides us with an ideal substrate to accurately characterize the band gap of ML-$Si_2Te_2$. We performed STM/STS measurements on a sample of ML-$Si_2Te_2$ grown on 1QL-$Sb_2Te_3$ (as illustrated in Figure 4a) and determined the local electronic structure (as shown in Figure 4b,c). First, we see that the four prominent peaks in the d$I$/d$V$ spectrum of ML-$Si_2Te_2$, as discussed before, are reproduced in Figure 4b. By zooming into a narrower energy window (Figure 4c), we can derive a band gap for 1QL-$Sb_2Te_3$ of 650 meV, as expected, and a band gap of 370 meV for ML-$Si_2Te_2$. The above excellent correspondence between the d$I$/d$V$ spectra and the DFT derived LDOS further demonstrates the experimental realization of ML-$Si_2Te_2$.

Having this ML-$Si_2Te_2$ material available, the next key issue is to identify the band topology of the obtained ML-$Si_2Te_2$ films grown on the $Sb_2Te_3$ substrate. It is well known that the topological insulating behavior of 2D TIs can be modulated by interfacial strain resulting from the presence of a substrate.[14,34,35] Considering the difference of the lattice constants between free-standing ML-Si$_2$Te$_2$ ($a_0$ = 3.88 Å)[20] and the obtained ML-Si$_2$Te$_2$ films ($a_{DFT}$ = 4.13 Å) on Sb$_2$Te$_3$, which results in a ~ 6.4% tensile strain, we calculate the band structure of free-standing ML-Si$_2$Te$_2$ films in dependence of external strain, ranging from -6% to 8% (see Figure S7). Figure 4d presents the global band gap of ML-Si$_2$Te$_2$ as a function of biaxial strain. Three different phases, metallic, QSH, and semiconducting, are revealed, which agrees well with the previous work.[21] The sketches in Figure 4d show the calculated three bands closest to the Fermi level (gray dashed line) at the Γ point with their parity eigenvalues indicate by "+" and "-". As shown, within the strain range from -3% to 2%, ML-Si$_2$Te$_2$ presents a QSH phase. However, once the tensile strain reaches 3%, a parity exchange between occupied and unoccupied bands is triggered at the Γ point, resulting in a phase transition from the QSH phase to a topologically trivial semiconducting phase. Our experimental data for ML-Si$_2$Te$_2$ grown on Sb$_2$Te$_3$ resides in this semiconducting regime, which is indicated by a red arrow in the phase diagram. In other words, the system can be driven into the nontrivial phase via reducing the strain by 4.4% (6.4% - 2.0%) using strain engineering. Figure 4e shows a set of tunneling spectra taken across a step edge between ML-Si$_2$Te$_2$ and 1QL-Sb$_2$Te$_3$ (individual d$I$/d$V$ curves are provided in Supporting Information, section S9). The absence of any kind of edge states provides further evidence for the strain modulated trivial band topology of ML-Si$_2$Te$_2$ by the Sb$_2$Te$_3$ substrate.

Based on our experimental and theoretical results, we propose two possible pathways for the realization of topological edge states in ML-Si$_2$Te$_2$ films. The first idea is based on mechanical exfoliation of ML-Si$_2$Te$_2$ films from Sb$_2$Te$_3$ substrates. The high coverage (95%) of our ML-Si$_2$Te$_2$ film as well as the weak vdW interaction between ML-Si$_2$Te$_2$ and the Sb$_2$Te$_3$ substrate indeed offers the possibility of mechanical exfoliation of large-scale ML-Si$_2$Te$_2$ films, which could provide a useful platform for investigations on free-standing ML-Si$_2$Te$_2$ films. The second idea is based on strain engineering of the epitaxial ML-$Si_2Te_2$ film. A typical way is to grow ML-$Si_2Te_2$ films on another substrate which has a lattice constant around 3.88 Å.

**CONCLUSION**

In conclusion, by combining STM/STS, XPS and DFT calculations, we provide compelling evidence, from surface morphology, composition and electronic structure, for the successful synthesis of ML-$Si_2Te_2$ films on $Sb_2Te_3$ substrates. A strain-dependent phase diagram of ML-$Si_2Te_2$ is obtained by DFT calculations, and the strain modulated trivial band topology of such films is revealed. The successful synthesis of high-quality ML-$Si_2Te_2$ films paves the way towards the potential room-temperature QSH phase of this material. In view of the compatibility with advanced Si processing technology, our experimental realization of this novel 2D material, $Si_2Te_2$, is expected to stimulate in-depth investigations of its potential for the application in optoelectronic devices and quantum computing.

**EXPERIMENTAL SECTION**

Details on the sample growth, STM/STS, XPS measurements and DFT calculations are provided in the Supporting Information.

**ACKNOWLEDGEMENTS**

The authors acknowledge W. Li, J. Gou, Q. Peng, and H. Kim for fruitful discussions. This work has been supported by the EU via the ERC Advanced Grant ADMIRE (No. 786020) and the DFG via the Cluster of Excellence "Advanced Imaging of Matter" (EXC 2056, Project ID 390715994). B.S. would like to thank for the financial support from the National Natural Science Foundation of China (No. 21973012), the "Qishan Scholar" Scientific Research Project of Fuzhou University, and the Natural Science Foundation of Fujian Province (Distinguish

Young Scientist No. 2021J06011). C.H. would like to thank the Alexander von Humboldt Foundation for support by a research fellowship. X.H thanks S. Samaddar for the help on the MATLAB code.

**CONFLICT OF INTEREST**

The authors declare no conflict of interest.

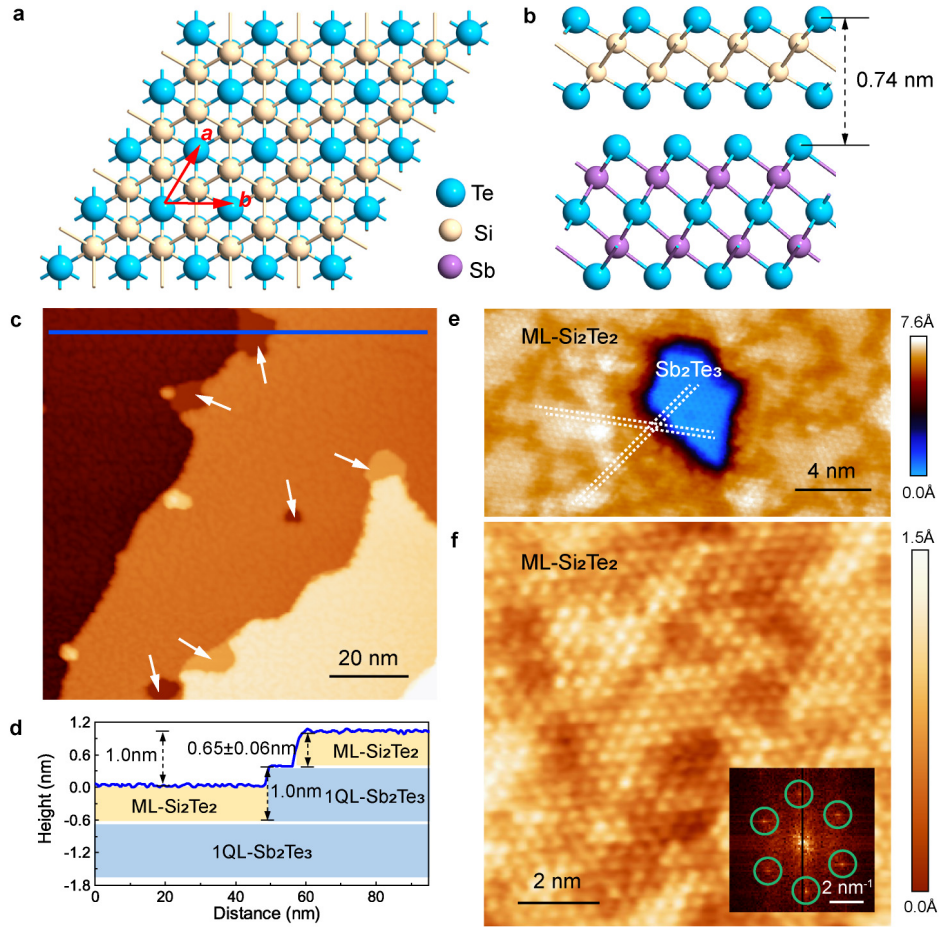

**Figure 1.** The morphology of ML-Si$_2$Te$_2$ on Sb$_2$Te$_3$. a,b) Schematic structural models of ML-Si$_2$Te$_2$ on 1QL-Sb$_2$Te$_3$. (a) Top view. (b) Side view. c) STM image ($V$ = -1.0 V, $I$ = 10 pA) of ML-Si$_2$Te$_2$ grown on a thick (>15 QL) Sb$_2$Te$_3$ film. The white arrows mark the small areas of the exposed Sb$_2$Te$_3$ substrate. d) Height profile, taken along the blue line in (c). Yellow and blue blocks represent ML-Si$_2$Te$_2$ and 1QL-Sb$_2$Te$_3$ films, respectively. Note that 1.0 nm is the step height of the Sb$_2$Te$_3$ substrate. e) High-resolution STM image ($V$ = -50 mV, $I$ = 500 pA) showing the surface lattice structures of both ML-Si$_2$Te$_2$ and the Sb$_2$Te$_3$ substrate simultaneously. White dashed lines are guides to the eye. f) Atomic resolution STM image ($V$ = -50 mV, $I$ = 5 nA) of Si$_2$Te$_2$. Inset: Fast Fourier transform pattern of (f). The six spots are marked by green circles.

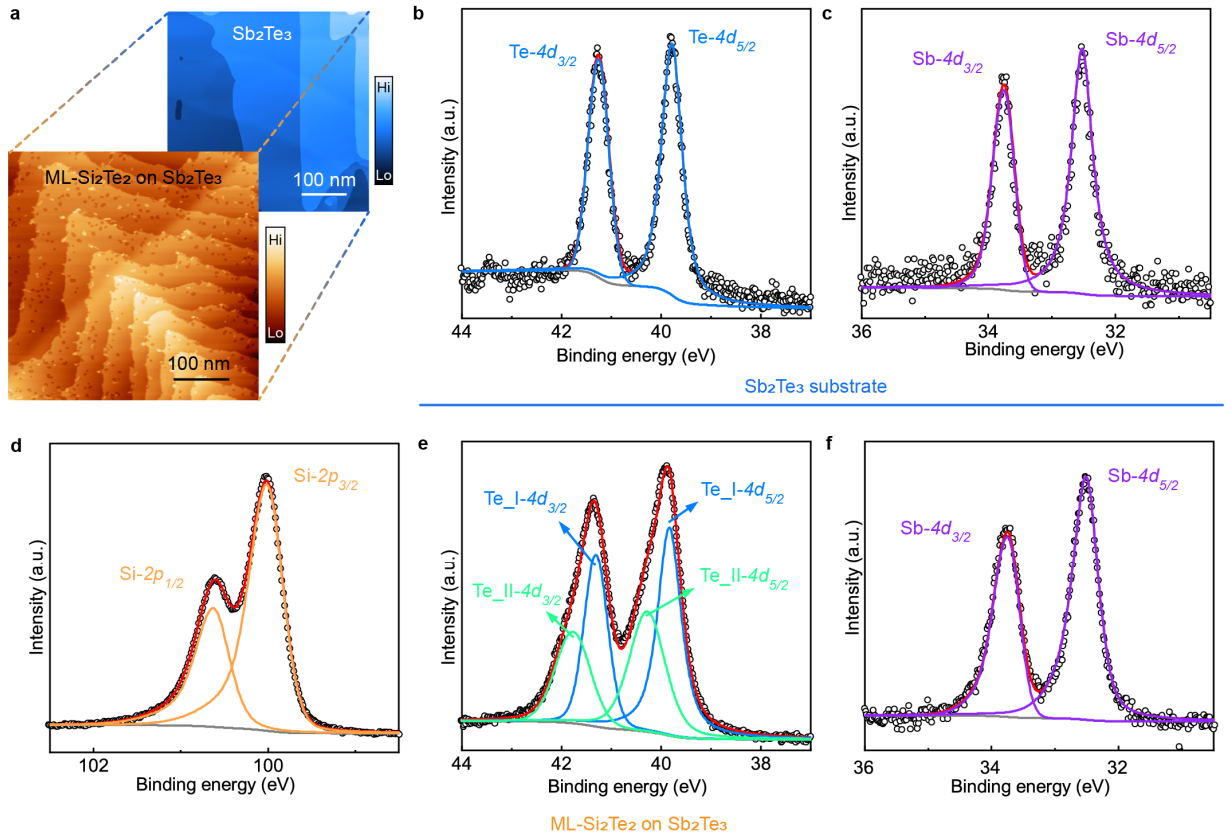

**Figure 2.** STM and XPS results of the Sb$_2$Te$_3$ film and the epitaxial ML-Si$_2$Te$_2$ on Sb$_2$Te$_3$. a) Large-area STM images taken on the thick Sb$_2$Te$_3$ film (upper right panel, $V$ = 2.0 V, $I$ = 5 pA) and ML-Si$_2$Te$_2$/Sb$_2$Te$_3$ (lower left panel, $V$ = -1.0 V, $I$ = 10 pA) samples before the XPS measurements. The small spots with darker contrast in the lower left panel correspond to the exposed Sb$_2$Te$_3$ film. b,c) XPS spectra of the Sb$_2$Te$_3$ substrate. (b) Te-4$d$ core levels. (c) Sb-4$d$ core levels. d-f) XPS spectra of ML-Si$_2$Te$_2$/Sb$_2$Te$_3$. (d) Si-2$p$ core levels from Si$_2$Te$_2$. (e) Te-4$d$ core levels. Two sets of peaks, Te_I-4$d$ and Te_II-4$d$, can be resolved. (f) Sb-4$d$ core level peaks from the Sb$_2$Te$_3$ film. The experimental data in (b-f) are displayed as black circles. A Shirley background (grey line) was subtracted before peak fitting. Blue and green lines represent two components of Te-4$d$, Te_I-4$d$ and Te_II-4$d$, violet lines represent Sb-4$d$ peaks, and orange lines represent Si-2$p$ peaks. Red curves correspond to the sum of the fitting lines.

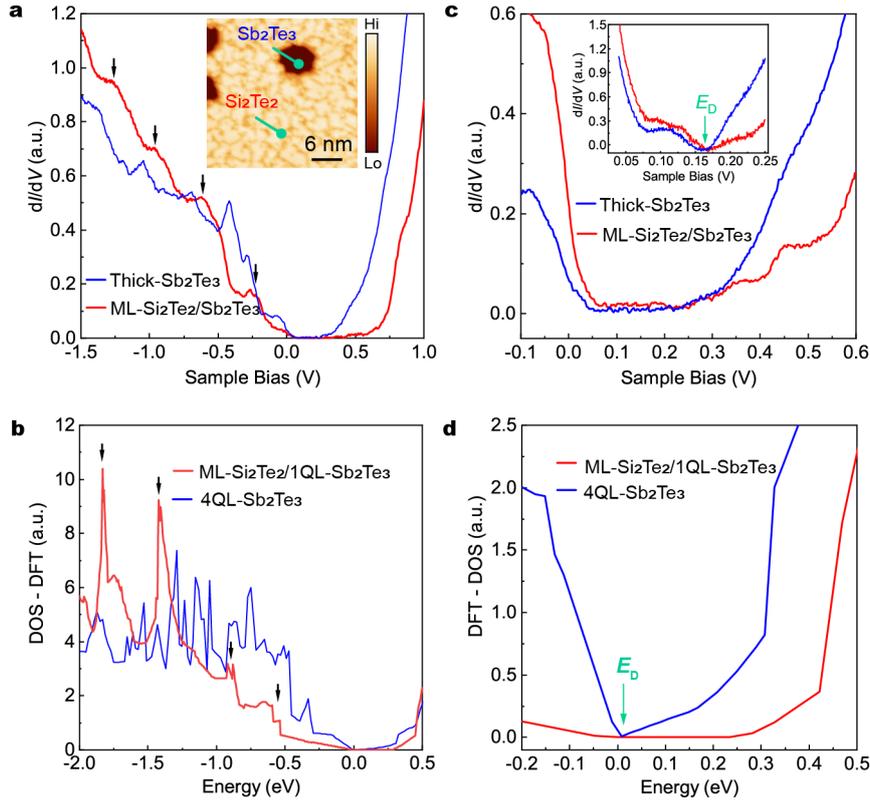

**Figure 3.** Tunneling spectroscopy on ML-Si$_2$Te$_2$/thick-Sb$_2$Te$_3$ (> 15 QL) films. a) Local d$I$/d$V$ spectra ($V_{stab}$ = -1.0 V, $I_{stab}$ = 100 pA, and $V_{mod}$ = 10 mV) measured on ML-Si$_2$Te$_2$ (red curve) and on the exposed Sb$_2$Te$_3$ film (blue curve) at the positions as marked in the STM image of the inset ($V$ = -1.0 V, $I$ = 10 pA). b) DFT derived DOS for the 4QL-Sb$_2$Te$_3$ and the ML-Si$_2$Te$_2$/1QL-Sb$_2$Te$_3$ heterostructure. Black arrows in (a) and (b) indicate four prominent peaks of the red curves below the Fermi level. c) Narrow bias range d$I$/d$V$ spectra ($V_{stab}$ = 0.8 V, $I_{stab}$ = 100 pA, and $V_{mod}$ = 10 mV). Inset: d$I$/d$V$ spectra measured with a much narrower tunneling gap ($V_{stab}$ = 0.25 V, $I_{stab}$ = 200 pA, and $V_{mod}$ = 1.0 mV). d) DFT derived DOS of ML-Si$_2$Te$_2$/1QL-Sb$_2$Te$_3$ (red curve) and 4QL-Sb$_2$Te$_3$ (blue curve) in a small energy range around the Fermi level. The green arrows in (c) and (d) mark the local minimum in the blue curve corresponding to the Dirac point. The d$I$/d$V$ spectra in (a) and (c) were recorded on samples with different Sb$_2$Te$_3$-thickness which results in different Fermi levels.

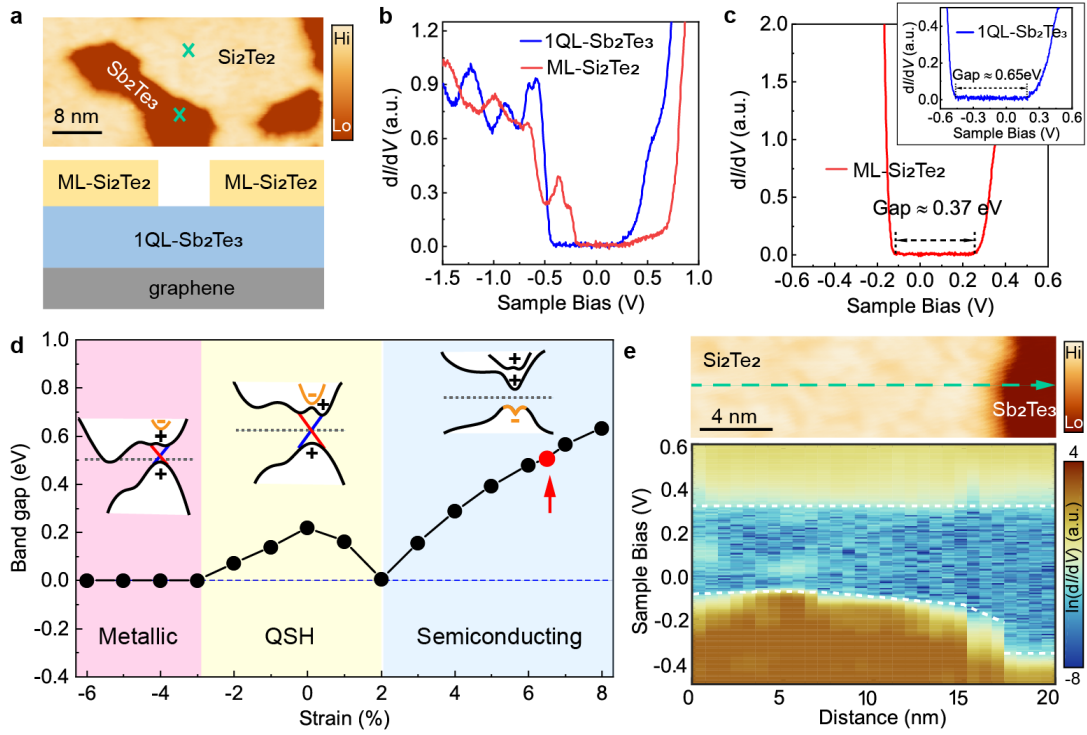

**Figure 4.** Band gap and strain-dependent phase diagram of ML-Si$_2$Te$_2$. a) STM image (upper panel) of ML-Si$_2$Te$_2$ grown on 1QL-Sb$_2$Te$_3$ ($V$ = 1.0 V, $I$ = 10 pA) and corresponding side view schematic (lower panel). b,c) d$I$/d$V$ spectra measured on 1QL-Sb$_2$Te$_3$ film (blue curve) and on ML-Si$_2$Te$_2$ (red curve) at the positions marked in (a). Measurement conditions of (b), (c) main panel, and (c) inset: $V_{stab}$ = -1.5 V, 0.6 V, and 1.0 V, $I_{stab}$ = 100 pA, $V_{mod}$ = 10 mV. d) Strain-dependent phase diagram with schematic band structures of ML-Si$_2$Te$_2$ with SOC effect calculated by using HSE06 hybrid functionals. The red arrow indicates the position where ML-Si$_2$Te$_2$ grown on Sb$_2$Te$_3$ is residing. e) Upper panel: STM image ($V$ = 1.0 V, $I$ = 20 pA) showing a step between ML-Si$_2$Te$_2$ and the 1QL-Sb$_2$Te$_3$ film. Lower panel: 2D color map of spatially dependent d$I$/d$V$ spectra ($V_{stab}$ = 0.7 V, $I_{stab}$ = 100 pA, and $V_{mod}$ = 5 mV) taken along the green dashed line in the upper panel on a one-dimensional grid with 0.67 nm spacing. White dashed lines are guides to the eye.